\documentclass[pra,aps,twocolumn]{revtex4}
\usepackage{graphicx}
\begin{document}
\title{Confinement of matter-wave solitons on top of a pedestal-shaped potential}
\author{K. K. Ismailov$^1$, B. B. Baizakov$^1$ and F. Kh. Abdullaev$^{1,2}$}
\affiliation{$^1$ Physical-Technical Institute, Uzbek Academy of
Sciences, 100084, Tashkent, Uzbekistan \\
$^2$ Department of Theoretical Physics, National University of
Uzbekistan, Tashkent 100174, Uzbekistan }

\begin{abstract}
Reflection of wave packets from downward potential steps and
attractive potentials, known as a quantum reflection, has been
explored for bright matter-wave solitons with the main emphasis on
the possibility to trap them on top of a pedestal-shaped potential.
In numerical simulations we observed that moving solitons return
from the borders of the potential and remain trapped for
sufficiently long time. The shuttle motion of the soliton is
accompanied by shedding some amount of matter at each reflection
from the borders of the trap, thus reducing its norm. The one- and
two- soliton configurations are considered. A discontinuous jump of
trajectories of colliding solitons has been discussed. The
time-shift observed in a step-like decay of the moving soliton's
norm in the two-soliton configuration is linked to the trajectory
jump phenomenon. The obtained results can be of interest for the
design of new soliton experiments with Bose-Einstein condensates.
\end{abstract}
\maketitle

{\it Introduction.} Matter-wave solitons are macroscopic objects
with particle-like properties, which can exhibit non-classical
behavior when they interact at low velocity with downward potential
steps and attractive potential wells. A classical particle would
certainly accelerate in these potentials, while quantum particle has
a finite probability to be turned back. Reflection from such
potentials occurs without reaching the classical turning point. In
this sense, quantum reflection is understood as a classically
forbidden reflection.

A counter-intuitive phenomenon of quantum reflection of bright
matter-wave solitons from attractive potential wells was
theoretically considered in Ref. \cite{lee2006} and later
experimentally observed in $^{85}$Rb Bose-Einstein condensate
\cite{marchant2016}. The subject was further elaborated in
subsequent studies for single solitons \cite{cornish2009, ernst2010}
and two-soliton bound states \cite{almarzoug2011}. Related
phenomenon for dark solitons was considered in \cite{sciacca2017}.

Significant quantum reflection occurs when the potential abruptly
changes over a spatial domain, much smaller than the width of the
wave packet. In ultra-cold quantum gases the de Broglie wavelength
of atoms can be considerably greater than the spatial region over
which the potential notably changes, thus suitable conditions for
quantum reflection of slowly moving atoms can
arise~\cite{pasquini2004,pasquini2006}. The confining property of
quantum-reflection traps in the form of a potential plateau is an
interesting subject~\cite{garrido2011}. The survival probability of
linear wave packets on top of a two-dimensional quantum reflection
trap was studied in \cite{madronero2007}. In experiments, a
pedestal-shaped trap for solitons can be created using
matter-wave-guides combined with red-detuned laser beams.

Nonlinear localized waves in pedestal-traps can exhibit novel
dynamical features, not observed with linear wave packets, in
particular with regard to their interactions and collisions. In the
early days of soliton research, the term ``soliton" was reserved
only for solutions of integrable nonlinear equations. Later the term
started to be used also for localized solutions of nonintegrable
equations possessing certain solitonic properties. True solitons
collide elastically and pass through one another preserving their
shape, amplitude, and velocity. One of the fundamental properties of
true solitons is that the trajectories of colliding solitons undergo
a discontinuous jump, which has been theoretically predicted
\cite{zakharov1972} and experimentally observed in optics
\cite{islam1991} and Bose-Einstein condensates \cite{nguen2014}.
Below we show that soliton collisions on top of a pedestal-shaped
potential can be a good platform for investigation of the
aforementioned ``trajectory jump" phenomenon, stemming from the
nonlinear interaction of colliding wave packets. A relevant study
for a soliton performing shuttle motion in a double-well potential
structure with an elevated floor and emitting linear waves was
reported in \cite{zegadlo2016}.

Our main objective in this work is to find out how long matter-wave
bright solitons can be held on top of a pedestal - shaped potential
while moving and experiencing repeated quantum reflections from its
borders. We also show that by comparing the step-like decay of the
amount of matter remaining on top of a pedestal-trap for one- and
two-soliton configurations one can estimate the magnitude of the
trajectory jump arising from soliton collisions. In the following
sections we describe the model and present the results of numerical
simulations of the governing Gross-Pitaevskii equation (GPE).

{\it Model.} In the experiments, the condensate is always held in
some external magnetic or optical trapping potential. If the
confinement in two of the spatial directions is much stronger than
in the third direction, the condensate acquires a highly elongated
cigar shaped form. The axial dynamics of the effectively one
dimensional condensate is described by the reduced GPE, represented
in normalized units as follows
\begin{equation}\label{gpe}
i\psi_t + \frac{1}{2}\psi_{xx}-V(x) \psi + |\psi|^2 \psi =0,
\end{equation}
where $\psi(x,t)$ is the mean field wave function of the condensate
and $V(x)$ is a plateau - shaped external potential
\begin{equation}\label{pot}
V(x)=\frac{V_0}{2} \left[{\rm th}\left(\frac{x+h}{w}\right) -   \right.
                   \left.{\rm th}\left(\frac{x-h}{w}\right) -
                   2\right],
\end{equation}
where $V_0$ is the drop in the potential, which takes place over
distance $w$.

The dimensionless variables in Eq. (\ref{gpe}) are related to
corresponding variables in physical units as follows $t \rightarrow
\omega_{\bot} t$, $x \rightarrow x/a_{\bot}$, $V(x) \rightarrow
V(x)/\hbar \omega_{\bot}$, $\psi \rightarrow a_{\bot}^{1/2}\,\psi$,
with $\omega_{\bot}$ being the frequency of radial confinement,
$a_{\bot}=\sqrt{\hbar/m\omega_{\bot}}$ is the harmonic oscillator
length, $m$ is the atomic mass.

The strength (drop) of the potential $V_0$, the length of the
plateau $2 h$ and width of the border region $w$ can be varied in
numerical simulations to find out suitable conditions for quantum
reflection. In absence of the external potential Eq. (\ref{gpe}) has
a fundamental one soliton solution, which will be used as initial
condition for Eq. (\ref{gpe})
\begin{equation}\label{soliton}
\psi(x,0)={\rm Sech}(x-x_0)\,e^{iv(x-x_0)},
\end{equation}
where $x_0$ is the initial position of the soliton and $v$ is the
initial velocity.

\begin{figure}[htb]
\centerline{{\quad a)} \hspace{3.5cm} b)}
\centerline{\includegraphics[width=4cm,height=4cm]{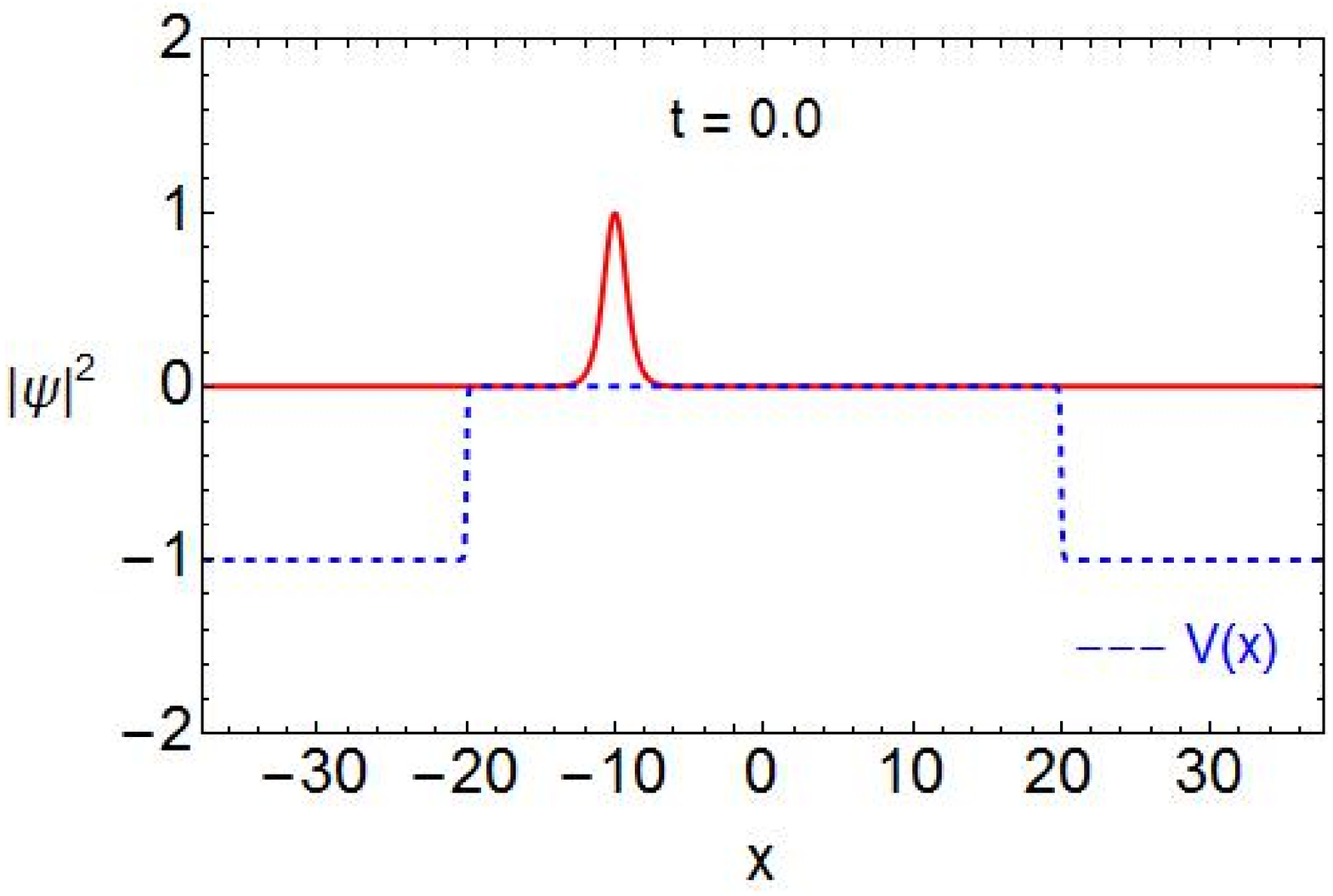}\quad
            \includegraphics[width=4cm,height=4cm]{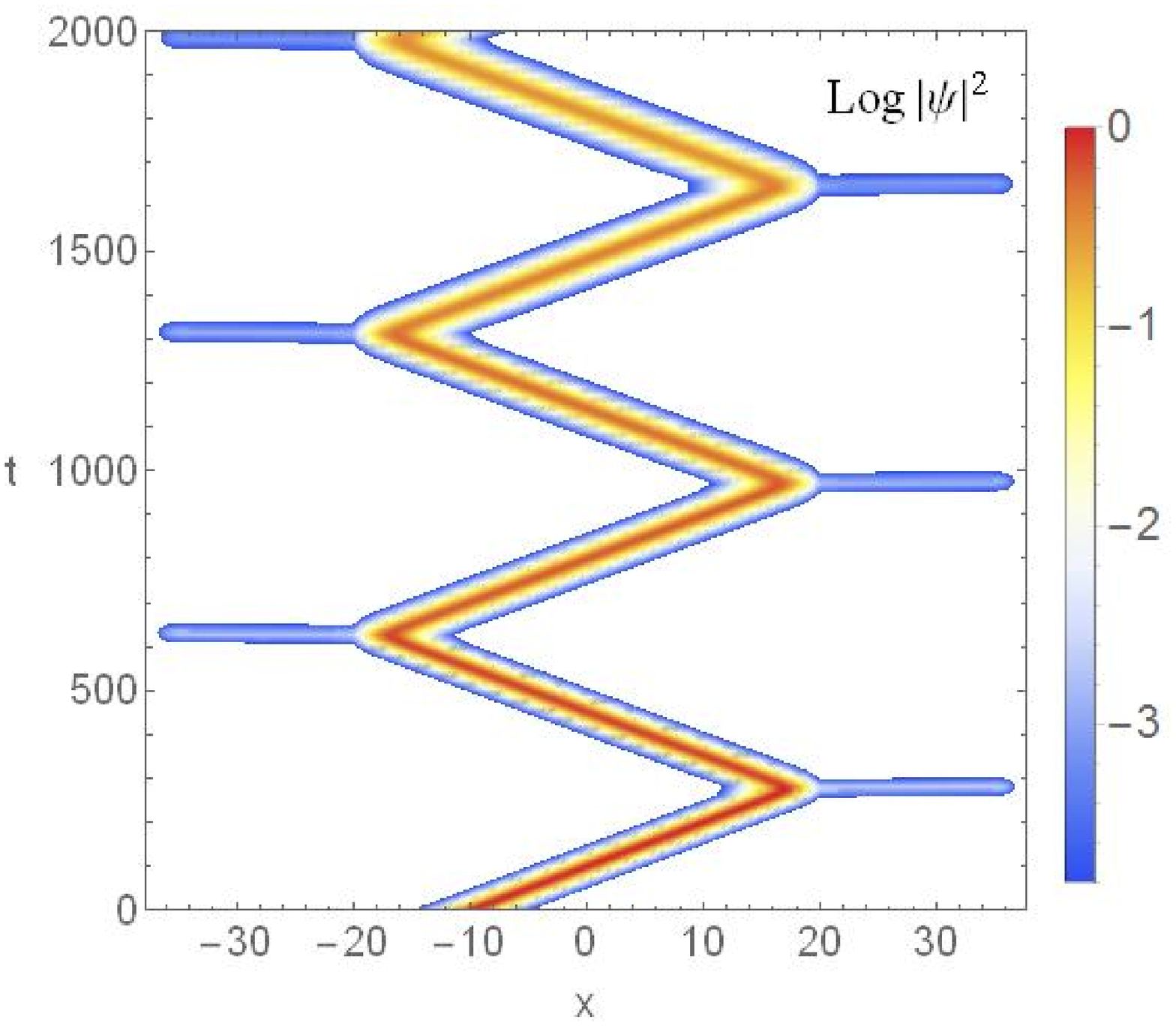}}
\caption{a) The sketch of the setup for confinement of a single
matter-wave soliton (red solid line) on top of a pedestal-shaped
potential $V(x)$, given by Eq. (\ref{pot}) for $V_0 = 10$, $h=20$,
$w=0.1$ (blue dashed line). For visual convenience we plot
$V(x)/V_0$. b) The density plot corresponding to numerical solution
of the GPE (\ref{gpe}) with initial condition Eq. (\ref{soliton})
and parameter values $x_0=-10$, $v=0.1$. The logarithmic scale for
$|\psi(x,t)|^2$ has been used to make the emission of linear waves
from the borders of the trap at $x=\pm \, h$ more visible.}
\label{fig1}
\end{figure}

{\it Numerical results for one-soliton configuration.} To verify if
a matter-wave soliton can be trapped for sufficiently long time on
top of a pedestal-shaped potential, we perform numerical simulation
of the GPE (\ref{gpe}). As initial condition we use the fundamental
soliton (\ref{soliton}), placed at $x=x_0=-h/2$, and set in motion
with some velocity. The result is shown in Fig. \ref{fig1}. When the
soliton is set in motion with sufficiently small velocity $v=0.1$,
it remains trapped on top of the potential, experiencing repeated
quantum reflections form its borders at $x = \pm h$.

\begin{figure}[htb]
\centerline{{\quad a)} \hspace{3.5cm} b)}
\centerline{\includegraphics[width=4cm,height=4cm]{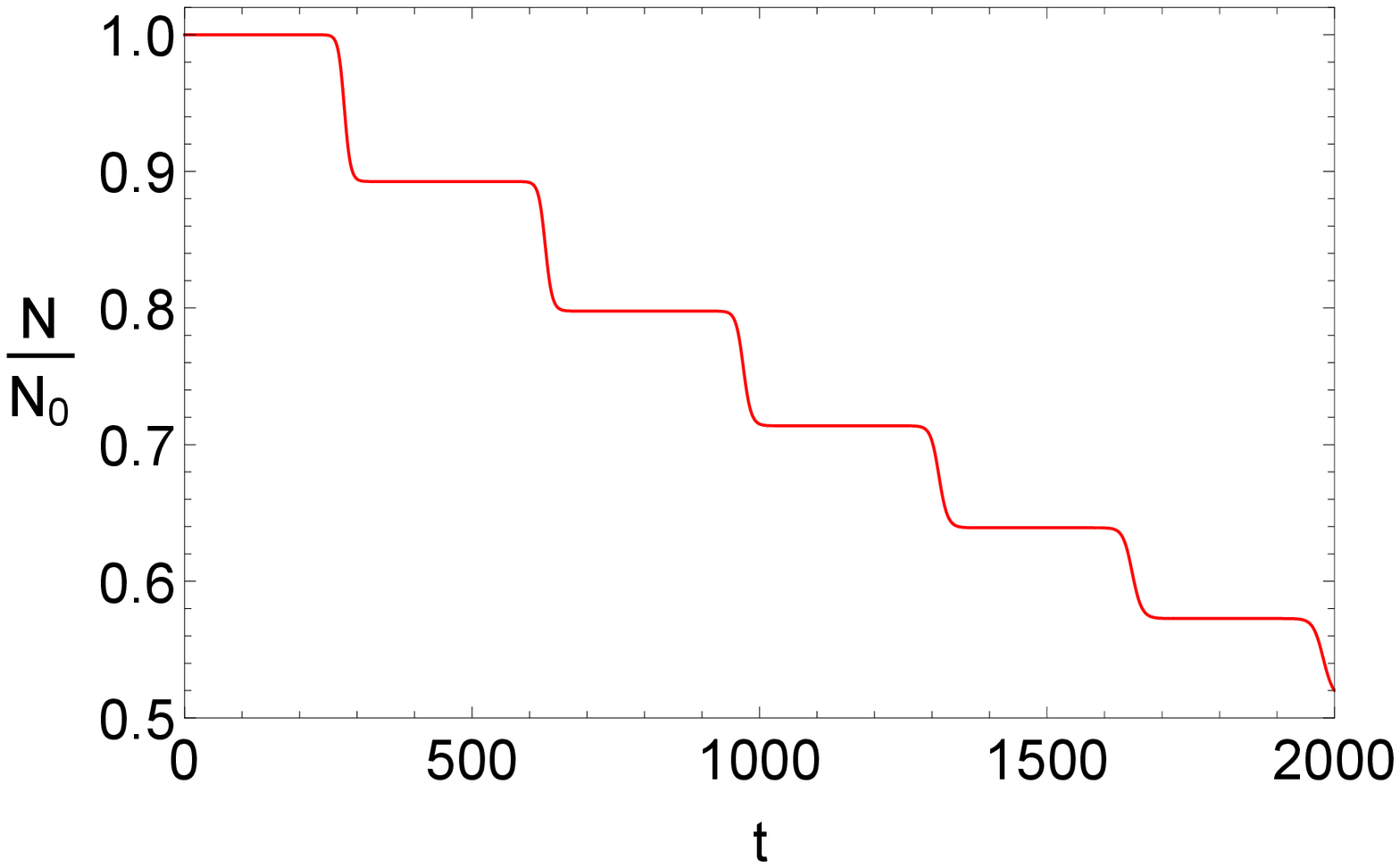}\quad
            \includegraphics[width=4cm,height=4cm]{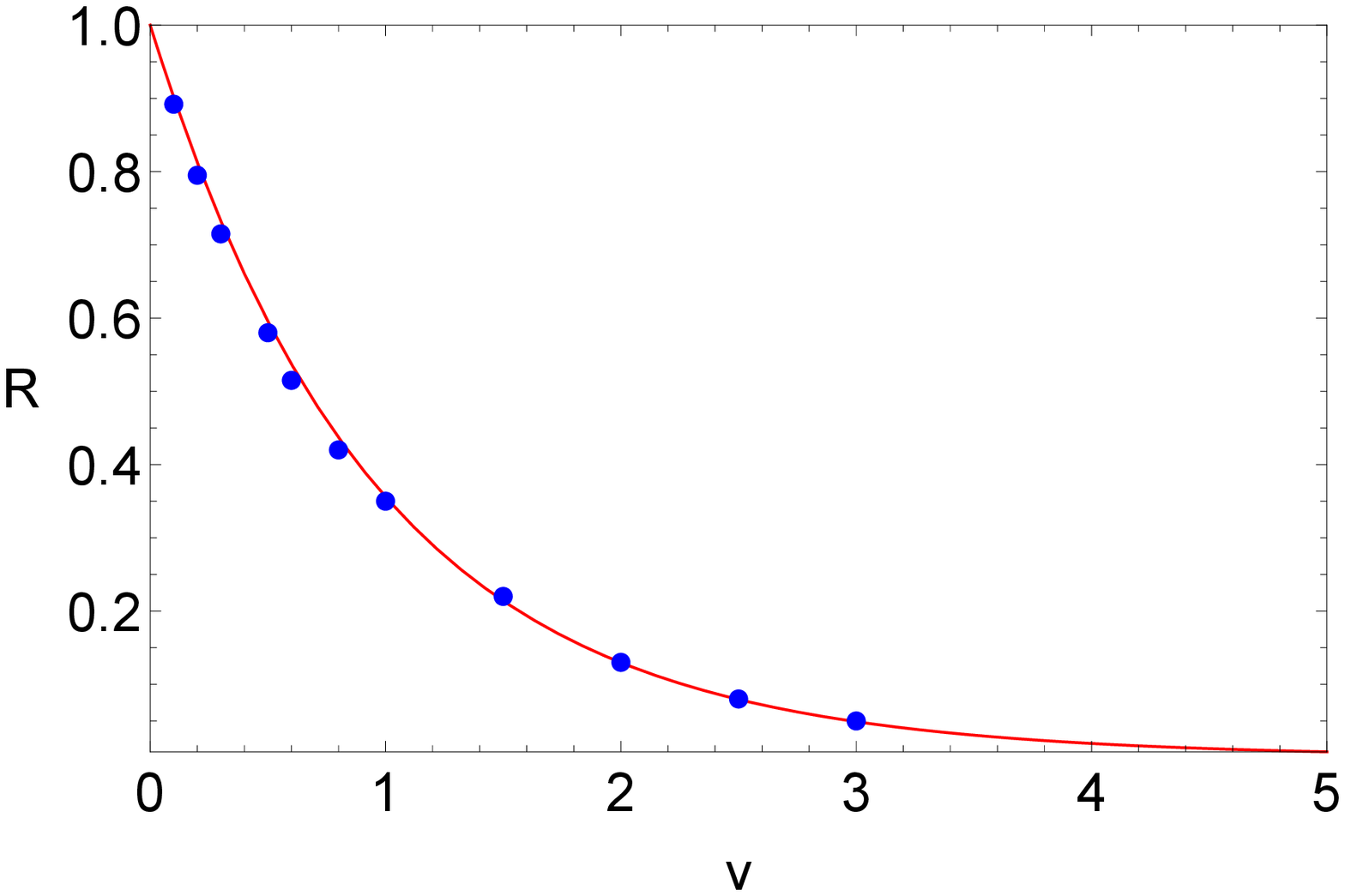}}
\caption{a) At each quantum reflection from the trap's border some
amount of matter is leaked out of the soliton, leading to stepwise
reduction of its norm $N$, proportional to the number of atoms in
the plateau region. b) The coefficient of reflection as a function
of the soliton velocity determined from the plane wave formula
(\ref{ref}) (red solid line) and numerical GPE simulations according
to Eq. (\ref{R}) (symbols).} \label{fig2}
\end{figure}

Each time the soliton is reflected from the trap's border some
amount of matter is transmitted to the adjacent region outside of
the plateau in the form of linear waves, where they can freely
propagate and eventually be absorbed in the domain boundaries. As a
result of this emission a stepwise reduction of the number of atoms
in the plateau-region $N(t)=\int_{-h}^{h} |\psi(x,t)|^2 dx$ takes
place, as illustrated in Fig. \ref{fig2}a. The absolute velocity of
the soliton performing shuttle motion on top of the pedestal-trap is
not affected by the reflections, while its norm diminishes.


From numerical simulations with different parameters of the
potential $V_0$, $w$ and initial the velocity $v$ we observe that
the trapping time strongly depends on the values of these
parameters. Namely, when the drop in the potential is large ($V_0
\gg 1$) and sharp ($w \ll 1$), the quantum reflection is
significant. This means longer time for the soliton to remain
trapped on top of the potential.

The stepwise time-dependence of the soliton's norm, shown in Fig.
\ref{fig2}a can be used to estimate the coefficient of quantum
reflection
\begin{equation}\label{R}
R = N_{i+1}/N_i,
\end{equation}
where $N_i, N_{i+1}$ are the norms (the amount of matter in the
plateau region) of the soliton before and after the reflection
correspondingly, $i$ is the step number in Fig.\ref{fig2}a starting
from the top. It turns out that the ratio $N_{i+1}/N_i$ is almost
the same for all $i=1,2,3...$, meaning that only the small-amplitude
tail of the soliton interacts with the trap borders, thus nonlinear
effects are insignificant.

For qualitative interpretation of obtained results we recall that in
the plane wave approximation, the reflection coefficient from a
negative potential step $V(x)=-(V_0/2)(1+{\rm th}(x/w))$ is given by
\cite{landau}
\begin{equation}\label{ref}
R_p = \left(\frac{{\rm Sinh}\left[\frac{\pi}{2}(k_2-k_1)w
\right]}{{\rm Sinh}\left[\frac{\pi}{2}(k_2+k_1)w \right]} \right)^2,
\end{equation}
where $k_1=v$, $k_2 = \sqrt{v^2 + 2 V_0}$. From Eq. (\ref{ref}) we
conclude that at fixed parameters of the potential $V_0$ and $w$,
the maximum reflection $R_p \rightarrow 1$ is observed for slowly
moving soliton $k_1 \rightarrow 0$, $k_2 \rightarrow \sqrt{2 V_0}$,
because in this case the numerator and denominator become equal. At
fixed velocity ($v$) and sharp potential drop ($w \ll 1$), the
coefficient of reflection is given by $R_p \rightarrow (k_2 -
k_1)^2/(k_2 + k_1)^2$~\cite{landau}.

It is relevant to mention, that although the Eq.~(\ref{ref}) is
derived for plane waves using time-independent Schr\"odiner
equation, our numerical simulations confirm its validity for moving
nonlinear localized wave packets, described by GPE (\ref{gpe}).
Since only the low-intensity tail of the soliton interacts with the
trap's borders, the plane wave formula (\ref{ref}) for the
reflection coefficient appears to be a good approximation. In Fig.
\ref{fig2}b we compare the prediction of Eq. (\ref{ref}) and the
corresponding results from GPE according to Eq. (\ref{R}). From the
obtained results we conclude that the matter-wave soliton can be
trapped on top of a pedestal-shaped potential for a long time
provided that its velocity is small enough to ensure significant
quantum reflection from the trap's borders.

\begin{figure}[htb]
\centerline{{\quad a)} \hspace{3.5cm} b)}
\centerline{\includegraphics[width=4cm,height=4cm]{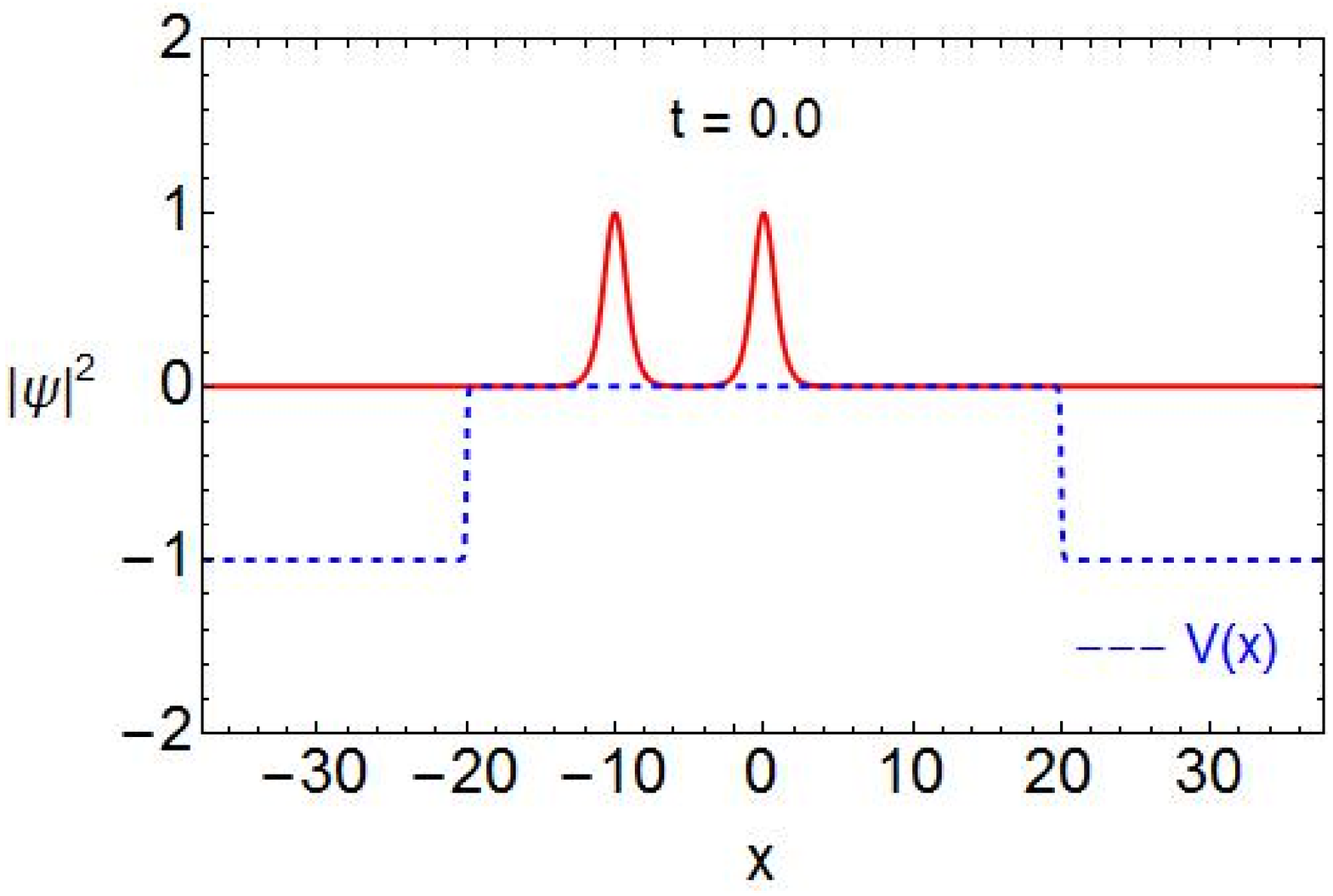}\qquad
            \includegraphics[width=4cm,height=4cm]{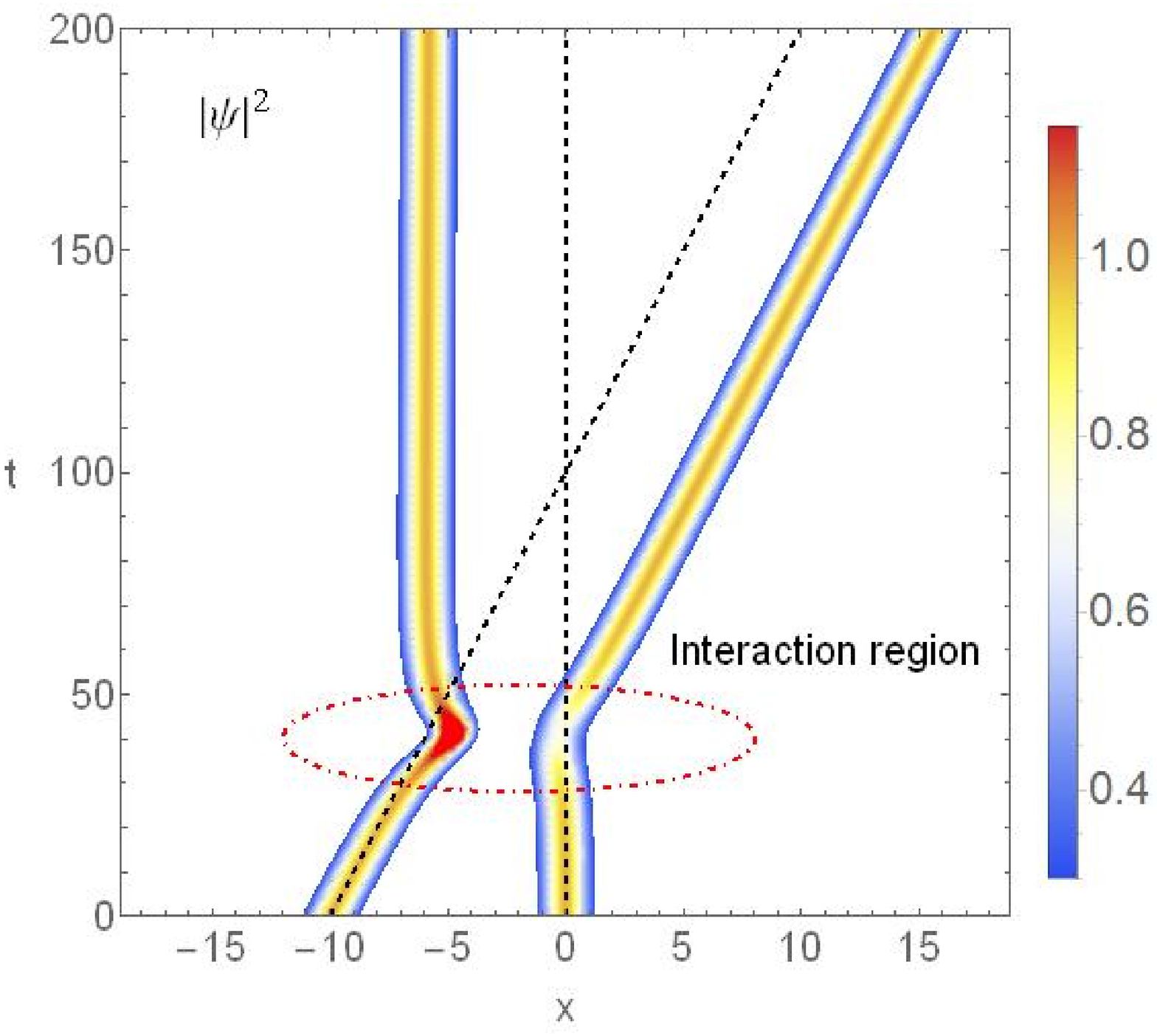}}
\caption{a) Schematic showing two solitons (red solid line)
positioned on top of a pedestal-shaped potential $V(x)$, given by
Eq. (\ref{pot}) for $V_0 = 10$, $h=20$, $w=0.1$ (blue dashed line).
For visual convenience we plot $V(x)/V_0$. b) The density plot
$|\psi(x,t)|^2$ corresponds to numerical solution of the GPE
(\ref{gpe}) with initial condition Eq. (\ref{soliton1}) and
parameter values $x_0=-h/2$, $v=0.1$. The moving soliton experiences
a discontinuous forward shift of the center-of-mass position, while
the quiescent central soliton shifts in backward direction. The
dashed lines show the soliton trajectories in the absence of
interaction.} \label{fig3}
\end{figure}

{\it Numerical results for two-soliton configuration.} Now we
consider the situation when two solitons are positioned on top of a
pedestal-shaped potential as shown in Fig. \ref{fig3}a.
The left soliton at $x = -h/2$  is set in motion towards the central
quiescent one at $x=0$ with a small velocity. The initial condition
for this setup is
\begin{equation}\label{soliton1}
\psi(x,0)= {\rm Sech}(x-x_0)\,e^{iv(x-x_0)} + {\rm Sech}(x).
\end{equation}
The two-soliton configuration is interesting because of the
intriguing phenomenon of trajectory jump, associated with soliton
collisions. The moving soliton experiences a discontinuous forward
shift of its trajectory after the collision with the quiescent
central soliton \cite{zakharov-book}
\begin{equation}\label{shift}
\Delta x = \frac{1}{\sqrt{2} A} {\rm
ln}\left|\frac{\lambda_1-\bar{\lambda}_2}{\lambda_1-\lambda_2}
\right|,
\end{equation}
where $A$ is the soliton amplitude. The complex parameter
$\lambda_1$ is linked to the moving soliton velocity and amplitude
as follows ${\rm Re}\lambda_1 = -v/4$, ${\rm Im}\lambda_1=A/2$. For
the quiescent soliton ${\rm Re}\lambda_2 = 0$, ${\rm
Im}\lambda_2=A/2$. The estimate (\ref{shift}) for the parameter
values $A=1$, $v=0.1$ yields $\Delta x \simeq 5.2$, in good
agreement with the GPE simulations shown in Fig. \ref{fig3}b. The
central quiescent soliton jumps in the backward direction for the
same distance and then remains immobile. Due to this phenomenon, the
moving soliton reaches the border of the trap for less time as
compared to the previously considered single soliton case. A similar
``time advance effect" was noticed also in the transmission of
solitons and two-soliton bound states through reflectionless
potentials \cite{almarzoug2011}.

\begin{figure}[htb]
\centerline{{\quad a)} \hspace{3.5cm} b)}
\centerline{\includegraphics[width=4cm,height=4cm]{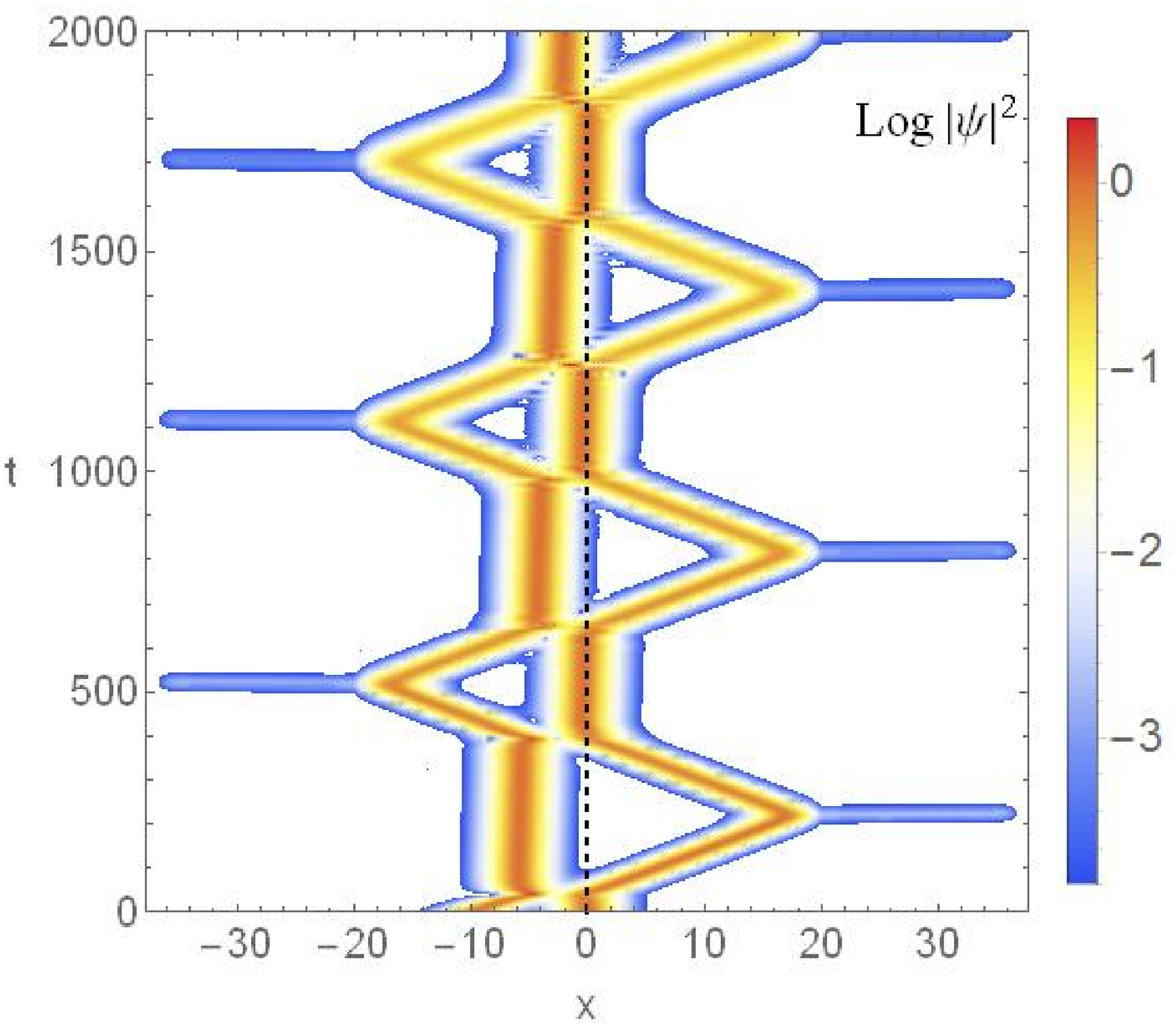}\qquad
            \includegraphics[width=4cm,height=4cm]{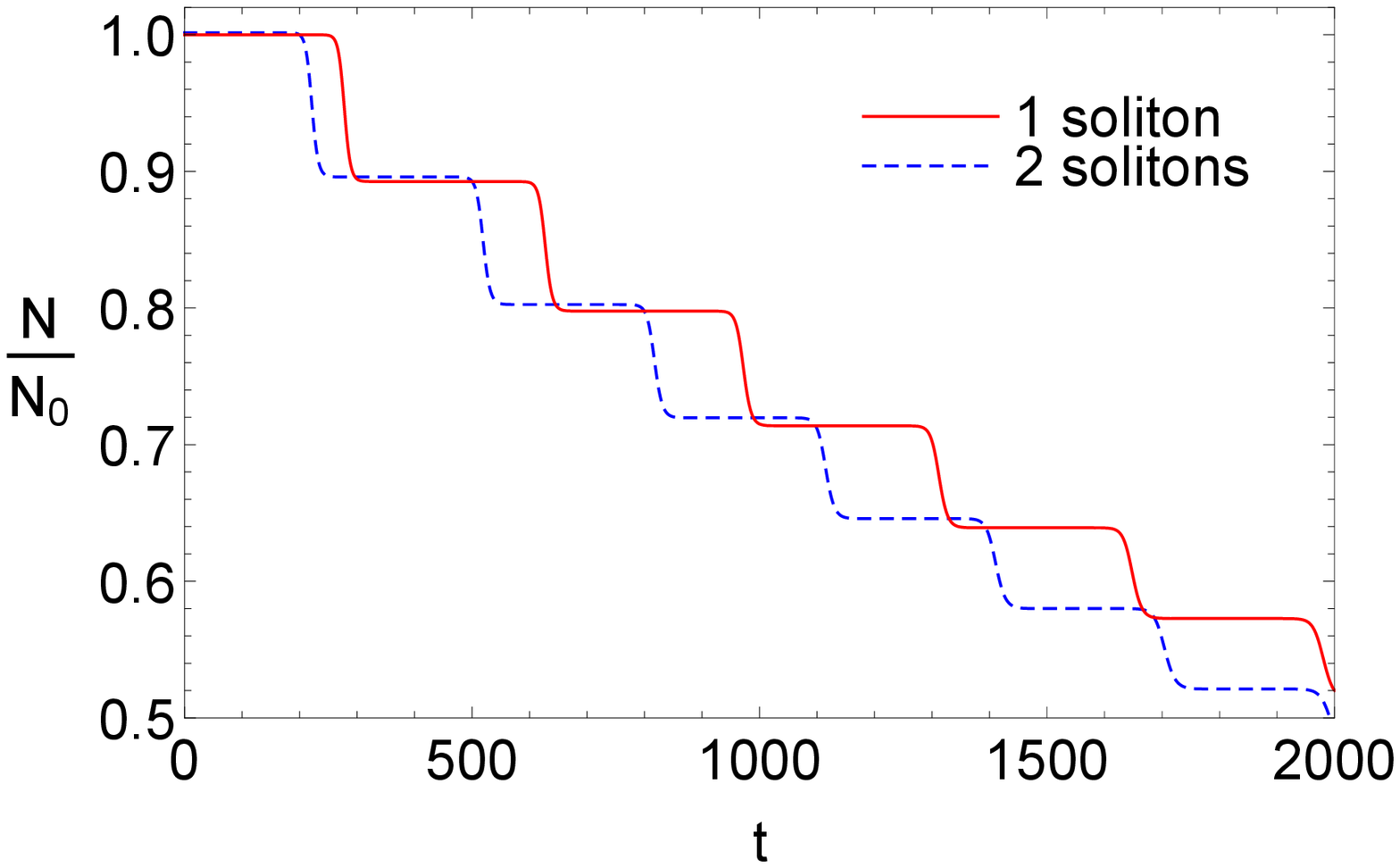}}
\caption{a) The density plot showing the repeated passage of the
moving soliton through the quiescent central soliton. The trajectory
shifts become smaller as the moving soliton's norm decays. The
dashed line shows the initial position of the central soliton. The
logarithmic scale is used to make the emission of linear waves by
the moving soliton at the trap borders more visible. b) The
magnitude of the trajectory shifts can be inferred from the stepwise
decay of the moving soliton's norm in a two-soliton configuration
relative to the one-soliton configuration. The parameter values are
the same as in previous figure.} \label{fig4}
\end{figure}

Figure \ref{fig4} illustrates how the trajectory jump of solitons
shows up in numerical simulations. The effect accumulates as the
moving soliton repeatedly passes through the quiescent central
soliton. Meantime, running into the borders of the pedestal-trap is
accompanied by simultaneous emission of small amplitude coherent
matter-waves by the moving soliton. Since the trajectory shift is
linked to the nonlinear interactions of solitons, its magnitude
decreases as the moving soliton's norm diminishes due to the
emission of linear waves at each reflection from the trap borders
(see Fig. \ref{fig4}a). By comparing the step-like decay of the
soliton's norm for the one- and two-solitons configurations, one can
estimate the magnitude of the trajectory jump phenomenon.
In particular, by comparing the extent of the 1st (top) step in Fig.
\ref{fig4}b one can see that the moving soliton in the two-soliton
configuration reaches the trap border earlier than in the one
soliton configuration by $\Delta t \sim 50$ time units. This
corresponds to shortening of the soliton path by $\Delta x = v
\Delta t = 5$, in agreement with the previous estimate with Eq.
(\ref{shift}). The subsequent soliton collisions give rise to
accumulation of the effect as is evident from comparing the solid
and dashed lines in Fig. \ref{fig4}b.

\begin{figure}[htb]
\centerline{{\quad a)} \hspace{3.5cm} b)}
\centerline{\includegraphics[width=4cm,height=4cm]{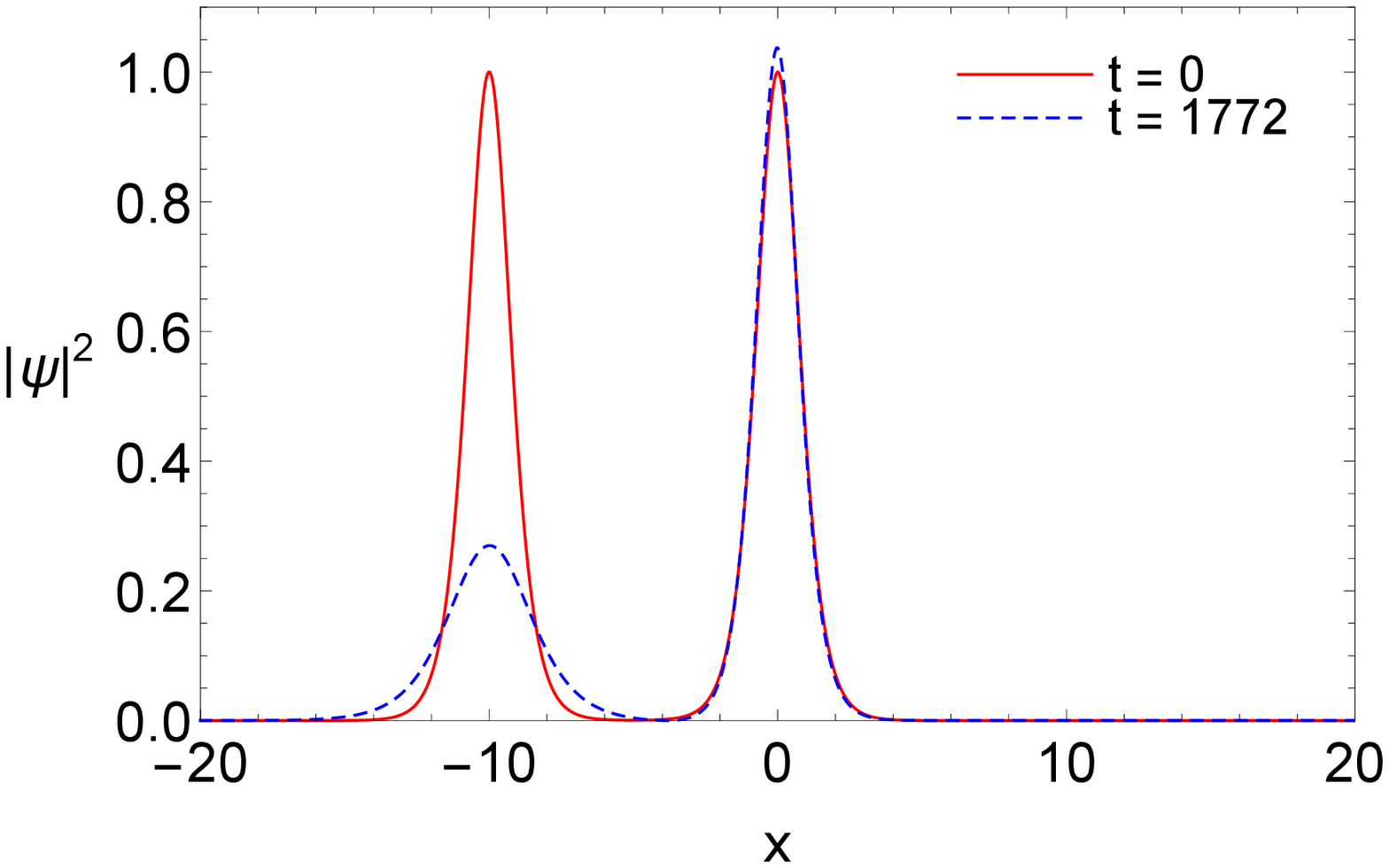}\quad
            \includegraphics[width=4cm,height=4cm]{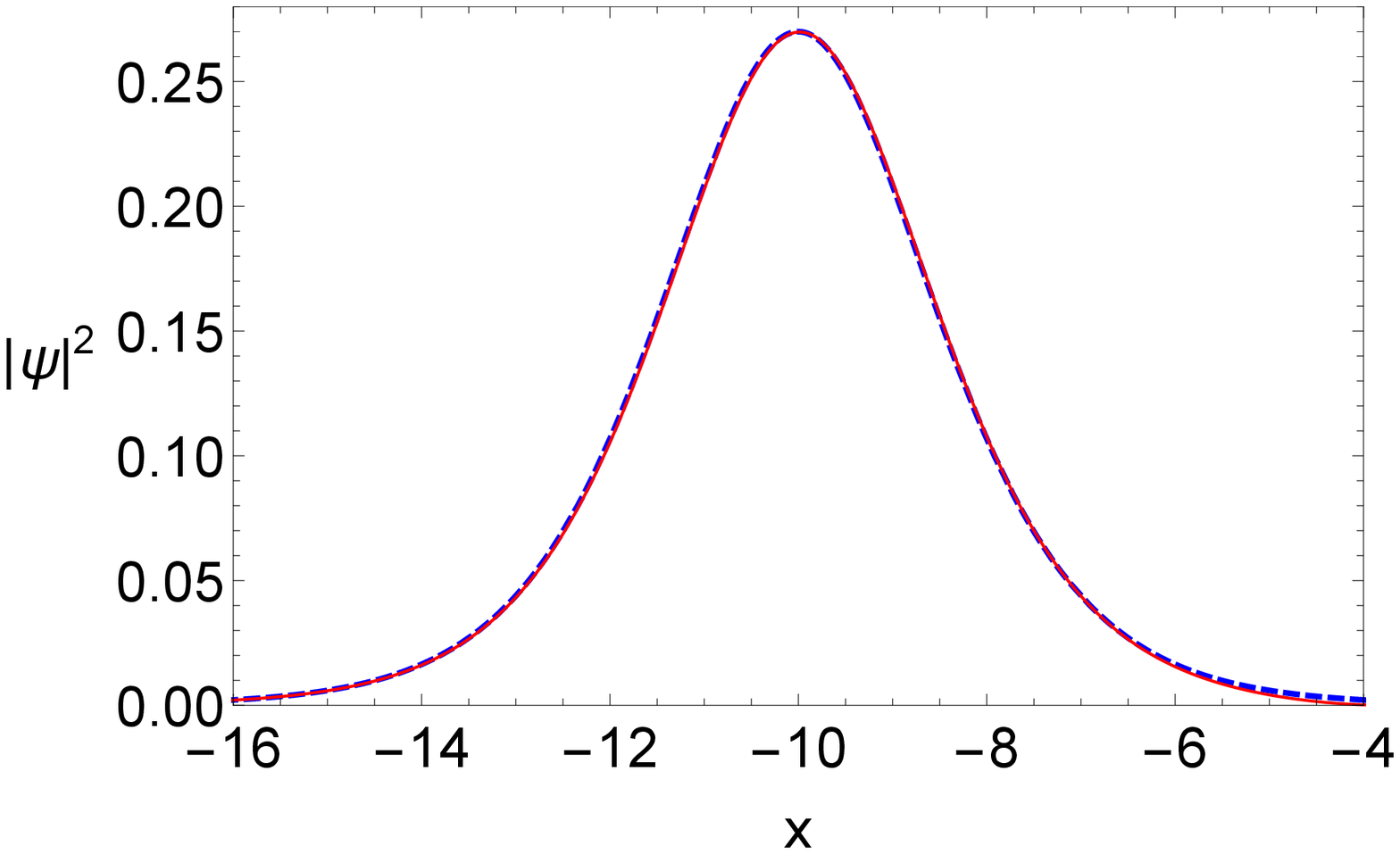}}
\caption{a) The shapes of the solitons at initial time ($t=0$) and
after six collisions ($t=1772$). The central quiescent soliton
preserves its initial shape and position, while the moving soliton
loses its norm due to emission of linear waves at the trap
boundaries. b) Inspection of the moving soliton's shape at $t=1772$
(blue dashed line) shows that it retains the form $A\, {\rm Sech}[A
(x-x_0)]$ (red solid line) with $A=0.52$, $x_0=-10$.} \label{fig5}
\end{figure}

An important question to be asked is whether the moving soliton
preserves its true soliton nature after repeated collisions with the
quiescent soliton. To clarify this issue we compare the shapes of
the moving soliton at an initial time and after six collisions with
the central soliton. As can be seen from Fig. \ref{fig5} the moving
soliton indeed preserves its original {\it Sech} form. This gives
the evidence that collisions were elastic inherent to integrable
systems. In the numerical simulations, the size of the integration
domain was sufficiently large ($L=24\ \pi$) and the velocity of the
soliton was relatively small ($v \sim 0.1$). These were necessary
conditions for the soliton to regain its Sech form after the
reflection from the trap boundaries and before the collision with
the central soliton.

{\it Conclusions.} We have demonstrated that moving matter-wave
solitons can be trapped on top of a pedestal-shaped potential for a
sufficiently long time. The mechanism behind this effect is the
quantum reflection, which occurs when a wave packet encounters with
downward potential steps or attractive potentials, rapidly varying
on a length scale much smaller than the width of the wave packet.
The intensity of linear waves, emitted to outside regions of the
trap potential, depends on the value of potential drop and sharpness
of the border. In the two-soliton configuration, we observed the
well-known phenomenon of trajectory jump in soliton collisions. As a
consequence of the last phenomenon, a time shift in the stepwise
decay of the soliton norm due to its quantum reflections on the trap
boundaries has been revealed. The phenomena considered in this work
can be observed in experiments with Bose-Einstein condensates, where
a high level of control over matter-wave solitons has been achieved
\cite{marchant2016,meznarsic2019}. Apart from basic scientific
interest, the results may have applications in the fields involving
coherent matter-waves.

\end{document}